**RESEARCH**  Open Access

# Augmented degree correction for bipartite networks with applications to recommender systems

Benjamin Leinwand[1*] and Vladas Pipiras[2]

*Correspondence:  
bleinwan@stevens.edu

[1] Department of Mathematical Sciences, Stevens Institute of Technology, 1 Castle Point Terrace, Hoboken, NJ 07030, USA  
[2] Department of Statistics and Operations Research, University of North Carolina at Chapel Hill, Hanes Hall, Chapel Hill, NC 27599, USA

## Abstract

In recommender systems, users rate items, and are subsequently served other product recommendations based on these ratings. Even though users usually rate a tiny percentage of the available items, the system tries to estimate unobserved preferences by finding similarities across users and across items. In this work, we treat the observed ratings data as partially observed, dense, weighted, bipartite networks. For a class of systems without outside information, we adapt an approach developed for dense, weighted networks to account for unobserved edges and the bipartite nature of the problem. The approach begins with clustering both users and items into communities, and locally estimates the patterns of ratings within each subnetwork induced by restricting attention to one community of users and one community of items community. The local fitting procedure relies on estimating local sociability parameters for every user and item, and selecting the function that determines the degree correction contours which best models the underlying data. We compare the performance of our proposed approach to existing methods on a simulated data set, as well as on a data set of joke ratings, examining model performance in both cases at differing levels of sparsity. On the joke ratings data set, our proposed model performs better than existing alternatives in relatively sparse settings, though other approaches achieve better results when more data is available. Collectively, the results indicate that despite struggling to pick up subtler signals, the proposed approach's recovery of large scale, coarse patterns may still be useful in practical settings where high sparsity is typical.

## Introduction

Recommender systems are widely used in industry to serve millions of users a variety of products and services. The goal is to uncover a similarity structure within and between sets of users and items that can be used to identify specific items for those users who are expected to be the target audience for those items. In general, user demographic features and item features like genre are available. In this work, however, we focus on the situation where the only information is users' ratings of items, where each user may rate a different subset of items. Ratings can be structured in a matrix with users on one axis, items on the other axis, and each cell containing the relevant user's rating of the relevant item. For services with large corpora and subscriber bases, most of the entries in this matrix will be missing. A dataset can be treated as a partially observed, dense, weighted, bipartite network.

The current state of the art in recommender systems uses variations of Graph Neural Networks (GNNs) (Zhou et al. 2020), and usually extensions of Graph Convolutional Networks (GCNs) (Hamilton et al. 2017a; Schlichtkrull et al. 2018). These techniques incorporate features of both users and items outside of the graph structure, like demographic or location information, as well as the edges on the network itself. While GNNs appear to succeed on a wide variety of tasks and graph structures, they also rely on message passing, which takes advantage of the topology of the observed network. In real-world systems, the existing edges may represent some signal, as the items with which users have chosen to interact may reflect some preference. For the specific task discussed in this paper, we focus on Neural Collaborative Filtering (NCF) (He et al. 2017) as a feasible alternative to our model. NCF learns an embedding for each user and item, and a function mapping these embeddings to estimated ratings. User defined hyperparameters include the embedding dimension of both users and items, the number of layers, the number of nodes in each layer, which nodes are connected, and the activation function of each edge, and the error metric. A good model is one which, given a user and an item, can approximately predict the value of the edge weight between the relevant user and item. Though we do not account for it here, a possible addition to the NCF framework is to include information about the users and items that is not contained in the network. For example, genre of an item may be known, and can be incorporated into the item's embedding. In the present context though, the only information available for edge estimation or any subtask leading to edge estimation (including user or item embedding), are the observed edges in the network.

The task of estimating unobserved edge weights is reminiscent of matrix completion, made famous by the Netflix Prize competition (Bennett and Lanning 2007). This topic has spawned its own literature, evidenced by Nguyen et al. (2019), Chen and Wang (2022) and the references therein. In general, these works assume the matrix is low rank, and try to find the matrix with the lowest rank that matches the observed entries, though Candès and Recht (2009) and Candès and Tao (2010) attempt to minimize the nuclear norm in place of the rank. As such, we also consider collaborative filtering techniques as potential alternatives to our approach. However, in this setting, as the functions used to generate the networks are nonlinear, this low rank assumption may not hold.

The method proposed in this paper can be seen as a nonlinear alternative approach to matrix completion. By segmenting users and items into communities, we can locally fit models to individual subnetworks, instead of the whole data set. We adapt a model for the creation and estimation of dense weighted networks that incorporates community structure and flexible connectivity patterns, known as "augmented degree correction," to this setting. Additionally, we introduce an algorithm that accounts for missing values and utilizes the bipartite structure of the network, performing community detection on the user nodes and item nodes simultaneously. We apply our methodology to a subset of the Jester Dataset (Jester Dataset 2001) and to simulated data.

The rest of this paper is structured as follows. In "Proposed model, measure, and algorithm" section describes the proposed model, a clustering measure, a community detection algorithm, and an estimation procedure suited to this setting. In "Alternative approaches" section details alternative modeling algorithms. In "Model evaluation" section defines the approach for measuring model performance. A simulation study is conducted in "Simulation Study" section, and the dataset and results are discussed in "Analysis of Jester dataset" section. Appendix 1 provides a deeper explanation of supplementary functions in our proposed community detection algorithm.

### Proposed model, measure, and algorithm

We approach the problem of modeling recommender systems by extending the work in Leinwand and Pipiras (2022) to bipartite networks, to account for missing values, and sometimes very few distinct values (e.g. when ratings only take on values in {1, 2, 3, 4, 5}). We shall discuss the relevant portions of the approach, but readers are referred to Leinwand and Pipiras (2022) for a more theoretical treatment of that model's motivation. The rationale for applying this approach mirrors the proposed benefits of that work, in particular to allow for community structure, nearly arbitrary distinct distributions of edge weights within and between communities, and "augmented degree correction" characterized by flexible "sociability" patterns (that is, the tendency for edges joining particular nodes to have persistently higher edge weights resulting in possibly non-convex contours demonstrating degree correction effects within the network) through the use of "*H*-functions."

The function $H_{ij} : (0, 1) \times (0, 1) \to (0, 1)$ is an *H*-function if:

- it is monotonic in both arguments
- $\int \int_{H(x,y) \leq z} dx dy = z$, for all $z \in (0, 1)$.

That is, if inputting two i.i.d. U(0, 1) random variables into an *H*-function, the output is distributed as one U(0, 1) random variable. *H*-functions have "positive association" if $H$ is non-decreasing in both arguments. Leinwand and Pipiras (2022) suggests a recipe for devising a subset of *H*-functions by taking cumulative distribution functions $F_1$ and $F_2$, as well as their convolution $F_{1,2}$, and setting

$$H(x, y) = F_{1,2}(F_1^{-1}(x) + F_2^{-1}(y)). \tag{1}$$

By construction, any function using this recipe satisfies the conditions of being a positive *H*-function The choice of $F_1$ and $F_2$ can define the concavity of the shape and tilt of contours of $H$ on (0, 1), allowing for a a variety of different maps. In the simulation study in "Simulation Study" section, we generate a network using *H*-functions with positive association, as well as those with negative association. $H(x, y)$ is an *H*-function with negative association if $H(1 - x, 1 - y)$ is an *H*-function with positive association.

In the *H*-Normal Nonlinear Sociability Model (NSM) in Leinwand and Pipiras (2022), nodes $u$ are assigned communities $i = 1, \ldots, K$, and "sociability" parameters $\Psi_u(k)$ with respect to all communities $k = 1, \ldots, K$, including their own, where all $\Psi$ values are assumed to be i.i.d. U(0, 1) random variables. The original model is not tailored to bipartite networks, so in the present situation, nodes $u$ will be considered users (or row nodes), while $v$ will be considered items (or column nodes). As a bipartite network, any edge must connect exactly one user with exactly one item, so we need only be concerned with modeling those kinds of edges. Consequently, $u$ may only belong to a user (or row) community $i$, and $v$ may only belong to an item (or column) community $j$. Each user $u$ is assigned "sociability" parameters $\Psi_u(j)$ with respect to each item (column) communities $\{j\}$, while each item $v$ is assigned "sociability" parameters $\Psi_v(i)$ with respect to each user (row) communities $\{i\}$. Each user/item pair of communities has their own possibly unique edge weight distribution $G_{ij}$, and a possibly unique *H*-function that maps the two incident nodes' $\Psi$ values to a single value, as well as a parameter $\sigma_{ij}$ which modulates how much the underlying sociability model is corrupted by random fluctuation. Edge weights $W_{uv}$ are generated by mapping both nodal parameters and the edge weight distribution into "standard normal" space. In particular, edge weights are drawn according to the following formula:

$$\Phi_1^{-1}(G_{ij}(W_{uv})) = \frac{1}{\sqrt{1 + \sigma_{ij}^2}} \Phi_1^{-1}(H_{ij}(\Psi_u(j), \Psi_v(i))) + \frac{\sigma_{ij}}{\sqrt{1 + \sigma_{ij}^2}} \epsilon_{uv}, \qquad (2)$$

where $\epsilon_{uv}$ is an idiosyncratic $\mathcal{N}(0, 1)$ random variable, and $\Phi_1$ is the CDF of a $\mathcal{N}(0, 1)$ distribution. By construction, the right-hand side of (2) is a $\mathcal{N}(0, 1)$ random variable, so $W_{uv}$ has distribution $G_{ij}$. Letting $\sigma_{ij} = 0$ would mean repeated sample networks from the same model would all be identical, while letting $\sigma_{ij} \to \infty$ would imply that each edge is a purely random draw from the relevant distribution $G_{ij}$, where the sociabilities of the incident nodes are immaterial.

**Community detection**

Due to the network structure, instead of estimating just one set of communities, users and items must each be partitioned into communities. Communities in this context are defined differently than in other work such as variants of the stochastic block model. Nodes in the same community $i$ all share common $G_{ij}$, $H_{ij}$ and $\sigma_{ij}$ for each particular corresponding community $j$. Instead of looking for homophily, patterns in edge weights can be used to cluster nodes into communities: users in the same row community $i$ share an "order of preferences" over items in a particular column community $j$, as reflected by persistently greater edge weights. Intuitively, a certain subset of users (e.g. those who love horror movies) may generally share an order of preferences over a certain subset of items (e.g. horror movies), while a different subset of users (e.g. those who hate horror movies), may share a completely different order of preferences over the same subset of items. This is different than describing item quality per se, as the different subsets of users may demonstrate diametrically opposite preferences, but it may broadly reflect latent features of the subsets of users and items (e.g. whether each user enjoys being scared, and the scariness of each movie). Perhaps less intuitively, though items do not have preferences, in this model, items in the same column community $j$ also share an "order of preferences" over users in a particular column community $i$, as reflected by persistently greater edge weights. In this sense, the user and item communities are interdependent, as fixing different partitions of users might induce different partitions of items, and vice versa. Let

$$d_i(v) = \sum_{u': u' \in i} W_{u'v}, v \in j,$$

be the local degree of node $v$ in community $j$ with respect to community $i$. Define a node-community correlation as

$$C_{ij}(u) = corr\{(d_i(v), W_{uv}) : v \in j\},$$

where missing values are ignored. The goal of a metric for community detection is to reward community structures where $C_{ij}$ values are consistently high and positive, while also preferring larger communities where possible so as to avoid overfitting too locally to only a few available edges.

To account for the bipartite structure and missing values, we must generalize the measure $L$ introduced in Leinwand and Pipiras (2022) to the context of recommender systems. In this case, let

$$\begin{aligned} L(\{\widehat{i}\},\{\widehat{j}\},\widehat{K_r},\widehat{K_c}) = \\ \sum_{\widehat{i}=1}^{\widehat{K_r}} \sum_{\widehat{j}=1}^{\widehat{K_c}} \left( \overline{C_{\widehat{ij}}(u)} \times \left(1 - \sqrt{SD(C_{\widehat{ij}}(u))}\right) + \overline{C_{\widehat{ji}}(v)} \times \left(1 - \sqrt{SD(C_{\widehat{ji}}(v))}\right) \right) \\ \times ((n_{\widehat{i}} - 2)(n_{\widehat{j}} - 2))_+ \times density(\widehat{i},\widehat{j}), \end{aligned} \quad (3)$$

where $\widehat{K_r}$ and $\widehat{K_c}$ respectively represent the numbers of estimated row (user) communities and column (item) communities, $n_{\widehat{i}}$ is the number of nodes in community $\widehat{i}$, and $density(\widehat{i},\widehat{j})$ represents the proportion of edges present in the subnetwork containing only row nodes in estimated row community $\widehat{i}$ and column nodes in estimated row community $\widehat{j}$. $\overline{C_{\widehat{ij}}(u)}$ and $SD(C_{\widehat{ij}}(u))$ refer to the average node-community correlation over nodes in community $i$ with respect to community $j$, and the standard deviation of those same values, respectively. In this form, $L$ values the row community $i$'s preference consistency over the column community $j$ equally to $j$'s preference consistency over $i$. The reasons to update the measure in Leinwand and Pipiras (2022) are that the matrix is no longer symmetric nor square, and that the missing values imply we should not merely account for the size of each subnetwork, but also for the subnetwork density; otherwise nodes with consistent observed preferences over only a few observed edges might have undue influence on the recovered community structures relative to those nodes with marginally less consistent preferences observed over a much greater set of incident edges. In a case where edges are not missing completely at random (MCAR), the very fact of greater (or lesser) interaction between users and items may itself be indicative of community behavior. This measure is the same as the measure in the original *H*-Normal NSM context from Leinwand and Pipiras (2022), except with respect to within community subnetworks. In this case, as the network is bipartite, there are no within community subnetworks.

The algorithm we propose to maximize this measure $L$ is described at a high level in "Community detection algorithm" section below, with additional detail in Appendix 1. It begins with a greedy algorithm to cluster singleton nodes into communities, which is a more complex version of Algorithm 2 in the supplement to Leinwand and Pipiras (2022). Instead of clustering both rows and columns simultaneously, this algorithm must alternate between rows and columns, agglomerating sets of singletons into larger communities. This initial agglomeration is especially difficult as $L$ depends on the interplay of row community detection and column community detection, but at the outset, there is no known community structure, and there is path dependency whereby the initial choices of which nodes to merge into communities can reverberate through the rest of the agglomeration step. That is to say, there is the least useful information available to the greedy algorithm at the beginning, but these earliest choices may have the greatest impact on the final recovered communities. As a result, the second step in the algorithm must be more complicated. It looks for nodes which have potentially been misclustered, and tries several different approaches to putting those nodes into alternative communities, and possibly combining communities. However, as community structure is dependent on other nodes in all communities, the algorithm alternates removing nodes from communities and recombining communities, continuing to iterate as long as the measure $L$ continues to increase after each cycle. At each cycle, the algorithm greedily seeks to maximize the increase in $L$, but as the possible proposed partitions are limited by the current configuration, the algorithm may attempt several possible partitions, with the aim of avoiding very local minima near the current estimated configuration.

**Community detection algorithm**

All row (user) and column (item) nodes begin in their own singleton communities, so the first stage is to alternate combining row communities and column communities, as long as the suggested change improves the measure $L$ in Equation (3). This is quite similar to Algorithm 2 from the supplemental technical appendix to Leinwand and Pipiras (2022), but has the extra wrinkle of having to account for both row nodes and column nodes, and includes adaptations of the attemptMerge and sweep procedures in Algorithms 5 and 6, as described in Appendix 1. When no more communities can be combined, this first part of the algorithm, agglomeration, shown in Algorithm 2, terminates. While this is intended to estimate a reasonable set of communities essentially from scratch, the issue is that row (or column) communities that are combined early on in agglomeration are not accounting for the column (or row) community structure that develops later. This is a feature, though, not a bug. Community detection is a partitioning problem, and NP-hard. As a means of getting any traction, agglomeration is designed only to combine communities, not to consider how to break them down more efficiently. Especially with a lot of missing values and no known community structure, it is to be expected that agglomeration will cluster many nodes into the wrong communities.

The goal of the second step, shown in Algorithm 3, is to identify and correct any clustering errors. The thorny issue that remains is one of interdependence. Even in perfect conditions with easily defined communities and $\sigma = 0$, if the column community assignments are wrong, a wrong partition of row nodes may outperform the true communities in the measure $L$. So while it's difficult to immediately reclassify misclassified nodes, there are heuristic approaches available, which can be used iteratively to get to better and better partitions. Those heuristics are used in Algorithms 7, 8, 9, and 10, which are deferred to Appendix 1, but at a higher level, Algorithm 3 starts with the results of agglomeration, and checks many alternative clusterings by using these heuristics, calling Algorithm 4. If the best performing alternative clustering achieves a higher $L$ than that of agglomeration, this new clustering is adopted and the process is repeated. If no heuristically induced clustering outperforms the current estimated communities, the algorithm takes the best performing result of the heuristics where all communities are of size > 2, and looks for alternative clusterings based on that structure. If any outperform the current clustering, the best is adopted and algorithm returns to the beginning. If not, as a last ditch effort, the model starts with the current clustering and attempts to combine each pair of row communities, then each pair of column communities. If the best performing change results in a higher $L$ than the current clustering, it is adopted and algorithm returns to the beginning. The whole purpose of all this machinery is to deal with the interdependence, where an early error in agglomeration may require several iterations to fix, yet to try to avoid being bogged down attempting every possible change.

**Algorithm 1** Greedy algorithm using *L*

---

    **Result:** $\{\widehat{i}\}, \{\widehat{j}\}$    // estimated community memberships for row (user) and column (item) nodes
    **Input:** $W$ // the observed network
**1** [rowLabels, columnLabels, rowCommunityAggregate, columnCommunityAggregate, L] = agglom($W, \{1, ..., n_r\}, \{1, ..., n_c\}$) // $n_r$ and $n_c$ are the number of rows (users) and columns (items), respectively

**2** $[\{\widehat{i}\}, \{\widehat{j}\}]$ = fixCommunities($W$, rowLabels, columnLabels, L)

---

**Algorithm 2** agglom

---

**Procedure** `agglom`($W$, *rowLabels*, *columnLabels*)
1. $q=1$;   rowLabels$^{(0)} = \vec{0}_{n_r}$;   rowLabels$^{(1)} = $ rowLabels
2. columnLabels$^{(0)} = \vec{0}_{n_c}$;   columnLabels$^{(1)} = $ columnLabels
3. rowCommunityAggregate $= W'$;   colCommunityAggregate $= W$
   // *rowCommunityAggregate* is a $\widehat{K_r} \times n_c$ matrix that treats each row community as a ``supernode,'' containing aggregate edge weights to each column node. *colCommunityAggregate* does the same for column communities.
4. **while** $rowLabels^{(q)} \neq rowLabels^{(q-1)} \,||\, columnLabels^{(q)} \neq columnLabels^{(q-1)}$ **do**
5.     $q$++
6.     rowLabels$^{(q)} = $ rowLabels$^{(q-1)}$;  columnLabels$^{(q)} = $ columnLabels$^{(q-1)}$
7.     rowLabelOrder = sort labels by decreasing column sum of rowCommunityAggregate
8.     colLabelOrder = sort labels by decreasing column sum of colCommunityAggregate
9.     **while** $\max(length(rowLabelOrder), length(colLabelOrder)) > 0$ **do**
10.         rowLabels$^{(q)}$, rowCommunityAggregate, rowLabelOrder = `attemptMergeBipartite`($W$, *rowCommunityAggregate*, *rowLabels$^{(q)}$*, *rowLabelOrder*, *columnLabels$^{(q)}$*)
11.         dequeue(rowLabelOrder(1))          // remove first entry
12.         columnLabels$^{(q)}$, columnCommunityAggregate, columnLabelOrder = `attemptMergeBipartite`($W'$, *columnCommunityAggregate*, *columnLabels$^{(q)}$*, *columnLabelOrder*, *rowLabels$^{(q)}$*)
13.         dequeue(colLabelOrder(1))          // remove first entry
        **end**
14.     **if** $rowLabels^{(q)} = rowLabels^{(q-1)} \,\&\&\, columnLabels^{(q)} = columnLabels^{(q-1)}$ **then**
            // if no communities are combined using *attemptMergeBipartite*, try to combine using *sweep*
15.         rowLabels$^{(q)}$, rowCommunityAggregate = `sweepBipartite`($W$, *rowCommunityAggregate*, *rowLabels$^{(q)}$*, *columnLabels$^{(q)}$*)
16.         columnLabels$^{(q)}$, columnCommunityAggregate = `sweepBipartite`($W'$, *columnCommunityAggregate*, *columnLabels$^{(q)}$*, *rowLabels$^{(q)}$*)
        **end**
    **end**
17. L = calculateMeasureL(W, rowLabels$^{(q)}$, columnLabels$^{(q)}$)
18. **return** rowLabels$^{(q)}$, columnLabels$^{(q)}$, rowCommunityAggregate, columnCommunityAggregate, L

**Algorithm 3** fixCommunities

---

**Procedure** fixCommunities(*W, rowLabels, columnLabels, L*)
1   oldMeasure = 0;        newMeasure = L
    **while** *oldMeasure < newMeasure* **do**
2      oldMeasure = newMeasure
3      newMeasure = getLargestOneStepL(*W, rowLabels, columnLabels*)
4      **if** *newMeasure(L)> oldMeasure* **then**
5        update rowLabels and columnLabels to those that produced newMeasure
     **end**
     **else**
6        largerGroups = newMeasure(largerGroups)
7        newMeasure = getLargestOneStepL(*W, largerGroups(rowLabels), largerGroups(columnLabels)*)
8        **if** *newMeasure(L) > oldMeasure* **then**
9          update rowLabels and columnLabels to those that produced newMeasure
       **end**

       **if** *newMeasure ≤ oldMeasure* **then**
10          Combine each pair of rowCommunities, and each pair of columnCommunities. If the best performing amalgamation produces a better *L* value than oldMeasure, update rowLabels, columnLabels, and newMeasure to those that produced this higher *L* value // As this requires $\binom{\widehat{K_r}}{2} + \binom{\widehat{K_c}}{2}$ comparisons, we only run this when all other, more efficient heuristics for adjusting community assignments fail
       **end**
     **end**
    **end**
11   **return** rowLabels, columnLabels

---

**Algorithm 4** getLargestOneStepL

---

**Procedure** getLargestOneStepL(*W, rowLabels, columnLabels*)
1  // Below, we write the assignment informally. Each
   assignment inherits the resulting rowLabels, columnLabels,
   rowCommunityAggregate, columnCommunityAggregate, and L, and
   possibly other results from the procedure on the right hand
   side. When feeding in results from one procedure to another,
   any necessary additional arguments are provided, but omitted
   here for brevity
   a = removeNodesA(*W, rowLabels, columnLabels*)
2  b = regroupNodesA(*a, banRemain = 0*)
3  c = regroupNodesA(*a, banRemain = 1*)
4  d = agglom(*a*); e = agglom(*b*); f = agglom(*c*)
5  g = removeNodesB(*W, rowLabels, columnLabels*)
6  h = regroupNodesB(*g, banRemain = 0*)
7  i = regroupNodesB(*g, banRemain = 1*)
8  j = agglom(*g*); k = agglom(*h*); l = agglom(*i*)
9  m = regroupNodesB(*a, banRemain = 0*); n = agglom(*m*)
10 **return** $\operatorname*{argmax}_{a,b,...,n}(L), \operatorname*{argmax}_{a,b,...,n}\{L : min(\text{community size}) > 2\}$ // The
   first list of returned variables contains all features of
   the best clustering according to the measure *L*. The second
   list of returned variables contains all features of the
   best clustering according to measure *L* where there are
   no communities of size 1 or 2, and is thus referred to as
   *largerGroups*. Both returned lists may be identical

---

**Model parameter estimation**

Once the communities have been estimated, the methodology proposed in Leinwand and Pipiras (2022) can be directly applied to learn model parameters. The estimation procedure can treat all subnetworks based on pairs of communities in this setting as "between community" subnetworks. In this section we will review that process, but as estimation is done at the subnetwork level without leakage from one subnetwork to another, all *i* and *j* subscripts are dropped for convenience. In short, *G* is estimated using a modified version of a nonparametric empirical CDF estimator, by simply ordering the edge weights and mapping this order to a uniform distribution. Ψ values are estimated by ordering the values $\sum_{v \in j} \Phi^{-1}(\widehat{G}(W_{uv}))$ within each community *i*. that is, by mapping the edge weights from their original distribution to a standard normal distribution, and ordering the nodes by their local degrees with respect to these transformed edge weights. The estimate of user *u*'s Ψ value with respect to community *j* is denoted by $\widehat{\Psi}_u^{(j)}$, while the estimate of item *v*'s Ψ value with respect to community *i* is denoted by $\widehat{\Psi}_v^{(i)}$. *H* and *σ* for each pair of communities is estimated by optimization using the following equation:

$$\{\widehat{H}, \widehat{\sigma}\} = \underset{H \in \mathcal{H}, \sigma \geq 0}{\operatorname{argmin}} \sum_{u \in i, v \in j} \left( \Phi_1^{-1}(\widehat{G}(W_{uv})) - \frac{1}{\sqrt{1+\sigma^2}} \Phi_1^{-1}(H(\widehat{\Psi}_u^{(j)}, \widehat{\Psi}_v^{(i)})) \right)^2. \quad (4)$$

where $\mathcal{H}$ is a set of $H$-functions. Finally, to estimate the edge weights in the network,

$$\widehat{W}_{uv} = \underset{w}{\operatorname{argmin}} \left| \Phi_1^{-1}(\widehat{G}(w)) - \frac{1}{\sqrt{1+\widehat{\sigma}^2}} \Phi_1^{-1}(\widehat{H}(\widehat{\Psi}_u^{(j)}, \widehat{\Psi}_v^{(i)})) \right|. \quad (5)$$

When not all edges are observed in the network, the network can be estimated by ignoring missing edges, and subsequently replacing the missing values with their estimate based on the incident nodes. The network can be iteratively re-estimated until the estimated edge weights converge. In this paper, we do not re-estimate until convergence, but instead use 10 iterations.

Graph Neural Networks (GNNs) utilize message passing to incorporate information from across the network via higher order relationships. Since this is a bipartite network, users never connect with other users, and because the networks of interest are assumed to be dense (even if edges are unobserved), every user is understood to be at a distance of at most two from any other user. In the estimation procedure, our approach accounts for not just neighbors, but all edges in the subnetwork. Each user's local sociability parameter is determined by its own local degree, but as the sociability parameter is based on a ranking, it also depends on the local degree of all other users in the subnetwork. All of these points hold true for items as well. Furthermore, the estimated $H$-function and $\sigma$ depend not only on the incident node's edges and estimated sociability parameters, but instead, the estimates rely on all other nodes and observed edges in the subnetwork. This implies that an estimated edge weight incident to a particular user depends not only on the user themselves and the item on the other side, but on how each second order neighbor in the user's community connects to every item in the item's community. In that sense, our approach can be understood as incorporating higher order relationships, albeit never needing to go beyond order two, because all nodes can be reached within two steps.

### Alternative approaches

Though Graph Neural Networks (GNNs) are a popular approach in recommender systems, they are not well suited to this task, as they implicitly rely on network topology as a source of information. That assumption is not unreasonable, as the missingness of an edge may reflect underlying preferences. For example some users may consciously avoid certain otherwise popular items, as they may not expect the item to suit their taste. However, the simulations in this paper operate under the assumption that edges are missing completely at random (MCAR). Under the MCAR assumption, incorporating network topology is spurious, as the presence or absence of an edge has no correspondence to item or user characteristics, even if the weight of that edge does. This also appears to be the case with the Jester dataset. As the underlying network is assumed to be dense, but merely partially observed, the topology of the network with all edges observed would only be a function of the edge *weights*, not adjacencies. Even so, we attempted to fit a model using message passing, but its poor performance did not warrant further investigation.

Neural Collaborative Filtering, on the other hand, appeared to show greater promise in data rich simulations. On the simulations below, we embedded both the users and items into 8-dimensional space, with 2 fully connected hidden layers made up of 64 and 32 nodes, where all activation functions are RELU, using MSE as the metric of interest. After some experimentation, these settings were chosen because they successfully captured the dynamics on our toy network when a large proportion of the edges are observed. Flexibility appears to be a strength of NCF, as this model was not purpose built for bipartite, dense, weighted networks. It is also not explicitly identifying communities at all, though those may be recoverable in the estimated embeddings.

We also tested several algorithms from the recommenderlab R package (Hahsler 2023) on our simulations and dataset. Specifically, we include: Item-based collaborative filtering (IBCF), User-based collaborative filtering (UBCF), Latent Factor Models using Singular Value Decomposition (SVD) with column-mean imputation, and the so-called Funk SVD using gradient descent. For more information on each of these algorithms, refer to Hahsler (2022) and the references therein. We do not include results from UBCF and SVD since they are always competitive with but inferior to IBCF results. In all cases, we train the algorithm on the entire training set using leave one out cross validation, and all results for these algorithms are based on the average of the resulting metric over for that algorithm over 10 runs. Other algorithms available in recommenderlab were ignored due to runtime, performance, or appropriateness for the task.

**Model evaluation**

Using our proposed pipeline from in "Proposed model, measure, and algorithm" section in the setting where only a single network is observed, comparing different models' performances is not straightforward. Instead, there are several options.

*Hold out edges* Treat a subset of the observed edges as missing values by withholding them during training. Communities can be estimated based on the observed edges, and the model can estimate the held out edges, which can be compared to the true, unseen ratings.

*Hold out users and items* Treat a subset of both users and items as missing, and train the model on the remaining data. After training, assign each held out user and item to the community that maximizes the measure in (3). Estimate the *observed* edges based on the model fit on the training set.

*Hold out users, items, and edges* Treat a subset of both users and items as missing, and train the model on the remaining data. After training, among the held out users and items, continue holding out a subset of edges, and assign each to the community that maximizes the measure in (3). Finally, estimate the missing edges using the model trained on only the training set.

Each of these options may be appropriate depending on the context. For example, when the sets of users and items are known and fixed, holding out edges alone may suffice. The second case may be helpful for understanding whether the community detection results generalize to new observations, as when new users join or items are added to the corpus. When new items and users are added to the dataset, but one can ask existing users about new items and new users about existing items, it may be best to hold out users, items, and edges. In all cases, assuming enough observations are present, the data can be split into training, validation, and test sets by holding out additional edges, users, and items as necessary. This may be useful for finding the best model, including choosing the best data transformation.

Unfortunately, only holding out edges alone works with the simplest versions of NCF, as the other options must still use newly observed edges to generate embeddings for previously unobserved users and items, thereby contaminating the estimates of the edge weights using the edges in the test set. This could be avoided if the embeddings were generated based on features of the users and items, as is the case for GraphSAGE (Hamilton et al. 2017b), rather than strictly based on ratings. This would also avoid the so-called "cold start" problem, when there are no edges observed for a new user or item, but outside information is available. Neither our model nor vanilla NCF is built to handle this problem, but conversely, GraphSAGE is not appropriate when the only available information is contained in the network itself. In this work, we hold out edges to compare models, but wish to explore the other methods in the future.

## Simulation study

The simulated dataset used for examining performance of the proposed algorithms is shown in the top left plot of Fig. 2. It consists of 3 item communities and 4 user communities. All communities have 73 nodes, and have nodes with $\Psi$ values equally spaced between .05 and .95, and all use the *H*-function as in (1) where $F_1$ and $F_2$ have the gamma density

$$f(x) = \begin{cases} \frac{-x^{-.5}e^x}{\Gamma(.5)}, & x < 0, \\ 0, & \text{otherwise,} \end{cases}$$

and $F_{1,2}$ is given by

$$F_{1,2}(z) = \begin{cases} e^z, & z < 0, \\ 1, & z \geq 0. \end{cases}$$

When the subnetwork of interest is on the diagonal, there is positive association, and the edge weights are distributed according to a U(0, 200) distribution. When the subnetwork of interest is on the off-diagonal, there is negative association, and the edge weights are distributed according to a U(0, 100) distribution. For all other subnetworks, there is negative association, and the edge weights are distributed according to a U(0, 50) distribution. When comparing different models, the whole dataset will not be observed during the training period. Instead, some subset of the dataset will be hidden, which can then be used as a test set on which to measure model performance. Of particular interest is the scenario when very few data points are available, as in many real world settings for recommender systems.

One shortcoming of this simulation is determining which edges will be observed, and which will be hidden. For our simulations, we conveniently assume edges are MCAR, which is unlikely to occur in real datasets, potentially due to different forms of selection bias. First, users know their own tastes and can often read reviews before trying an item, so higher valued edges may be more likely to be observed. At the same time, users may be more likely to leave a rating for items which elicited strong feelings, whether they were positive or negative. There may also be some interaction with a recommender system algorithm, where users believe their ratings will shape future recommendations. The combination of these behaviors would expect to result in a relatively large number of extreme valued edges for many users, but relatively few edges with moderate values. It may be desirable to sample edges in a fashion consistent with these behaviors in future work, as well as to use real datasets that are known to feature these patterns of omissions.

"Our proposed method" refers to the pipeline described in "Proposed Model, Measure, and Algorithm" section, including both community detection and estimation (though we also include the case where communities are known in advance). When the estimation procedure observes and estimates a network with missing values, the unobserved values are replaced with their estimates, then the network is re-estimated. Each network (except the fully observed network) is estimated a total of 10 times, though not necessarily until convergence, and the tenth estimate is treated as the final estimate. It may be of interest to feed the network with estimated missing edges back into the community detection algorithm, though this was not done. For all NCF results, we fit the model described in "Alternative approaches" section for 1000 epochs and a batch size of 2048 in all epochs.

**Simulation performance and comparison**

A total of 21 networks were generated. The first network is the full network, with no edges missing, so the only meaningful performance reported is that of the NCF on the training set, if only to demonstrate that the choice of hyperparameters can suitably capture the network dynamics. The rest of the networks are 4 sets of 5 networks that have, respectively, 50%, 70%, 80%, and 90% of the edges deleted at random. This model has $\sigma = 0$, though the interaction between missing values and the error magnitude may be worthy of investigation.

Our proposed method has two modules: community detection and estimation. The estimated communities will impact the estimation performance, so we seek to examine the performance of both modules, as well as the overall performance of the combined methodology as the number of missing edges increases. The left plot in Fig. 1 shows the Normalized Mutual Information (NMI) of the community memberships and their estimates. The community detection algorithm performs perfectly in all cases when at most 70% of the edges are missing. The model also performs well when 80% of the edges are missing. However, community detection fails when 90% of edges are missing. We suspect the issue is not due to the proportion of missing values, but rather the lack of observed edges incident to each node. If we kept the same sparsity but increased the size of the network to observe some threshold number of total edges incident to each node, we would expect the community detection to succeed. We have begun to test this hypothesis, first by removing every other user and item node from a network missing 50% of its edges. In this case, community detection failed. Next, we took a network missing 80% of its edges where community detection failed to recover the true communities, and duplicated all the nodes, thus creating a network with 4 times as many edges, but no new information. In this case, true communities were recovered.

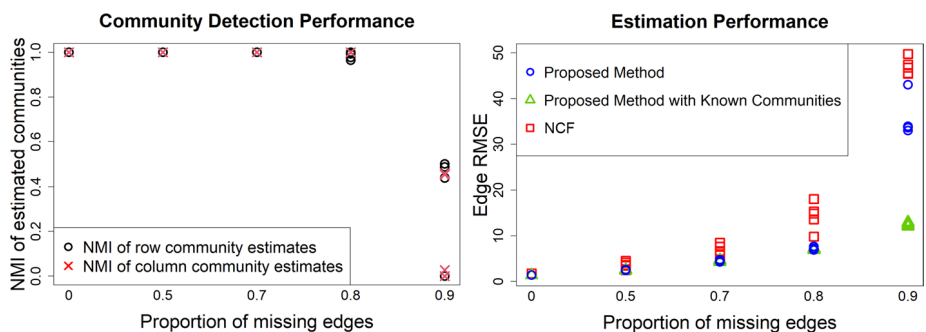

**Fig. 1** Community detection performance of our method, and comparing estimation performance with NCF

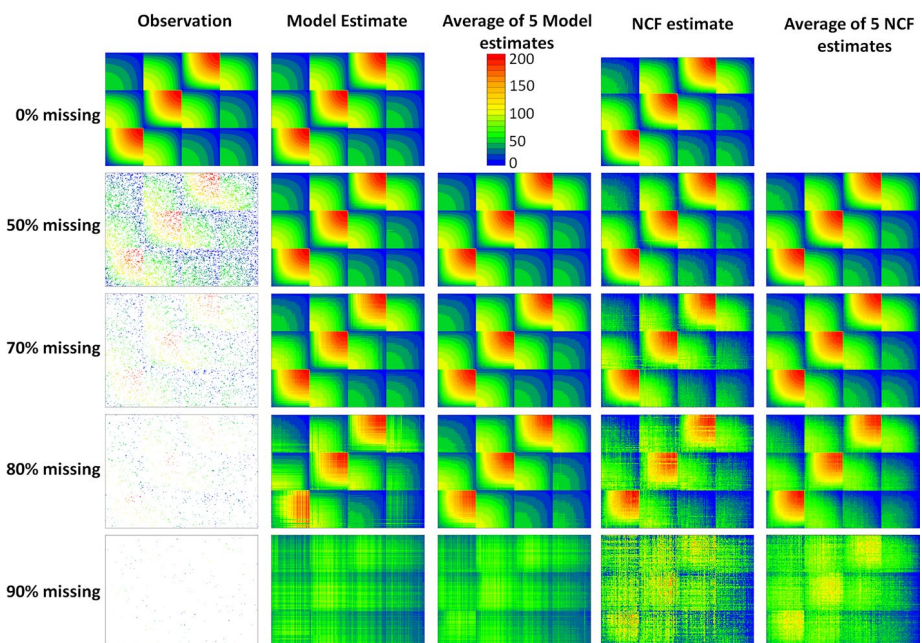

**Fig. 2** By column from left to right: observed networks, our model's estimates, the average of 5 model estimates, NCF estimates, and the average of 5 NCF estimates

The left column of Fig. 2 shows the network the model observes, where white spaces represent unobserved edges. The network in the first row contains all edges, while the next 4 rows are missing 50%, 70%, 80%, and 90% of the edges, respectively. In the second column from the left, the results of the estimation procedure of our proposed method. In the middle column, the average of 5 estimated networks, where each estimate is based on a different observed network which contains the same proportion of missing edges as the observed network in that row. In cases with many missing edges, the model may expect too much from nodes that randomly are not missing their largest edges. Furthermore, the model may not have a large number of different edge weights to choose from within each subnetwork, potentially resulting in resolution loss. Even with much of the data missing, the model recovers the broader pattern, though edges incident to specific nodes may be overestimated or underestimated, resulting in disordered or "plaid" looking patterns. However, when community detection fails, so too does estimation, as the pattern of preferences is more muted due to the mixing of positive and negative associations for nodes across the network.

The NCF estimates appear to degrade even more sharply than using our proposed method. The right plot in Fig. 1 shows the RMSE of the estimates for the 21 different networks using the estimates from NCF, our proposed method including community detection, and our proposed method with known communities. While all fit the fully observed network nearly perfectly, a divergence begins when 70% of the data is missing, and the 3 methods are clearly in separate clusters when 90% of the data is missing, with NCF performing the worst. This plot shows the RMSE of edges in the test set, but when the whole network is observed, there is no test set, in which case the RMSE on the training set is reported. The right two columns of Fig. 2 depict the estimates and averages from the fitted NCFs for different levels of sparsity, which look somewhat worse than our method's estimates.

The analysis indicates that community detection is pivotal. If the true communities are recovered, the model can broadly reconstruct connectivity patterns, even when most edges are missing, as illustrated in Fig. 3. The left plot of that figure displays the estimated 90% missing network shown in Fig. 2 using the true communities. The right plot of Fig. 3 represents the average of 5 estimates using the true communities, rather than the estimated communities. When there are few nodes and many edges are missing, though, community detection may remain a difficult problem, simply because there isn't enough information in the dataset. Recycling nodes to increase absolute coverage appears to be a promising option, though when error is included (that is, when $\sigma > 0$), that error will also be amplified.

Table 1 includes detailed results from the simulation. Our proposed method outperforms other algorithms when at most 80% of the edges are missing, and the NCF is competitive when most 70% of the edges are missing. A few interesting things occur when the network gets very sparse, though. First, poor community detection means other approaches are preferred when 90% of the edges missing. Even more surprisingly, while the Funk SVD algorithm performs poorly when at least 20% of the network is observed, it is by far the best performer with only 10% of the data, the only algorithm to improve with less information.

**Analysis of Jester dataset**

Using the Jester Data set from Goldberg et al. (2001) of joke ratings from 24,983 users who have rated at least 36 out of 100 jokes, we include only those users who rated exactly 74 jokes. This was chosen to get a feasibly large dataset, and one with a reasonable amount of missing data (the network is chosen to be 26% missing). This leaves 626 users and 100 jokes. The jokes do not appear to be MCAR, but instead many jokes are far more frequently rated than others, likely due to ordering effects. Ratings run from −10 to 10. The distribution of ratings for either users or items shouldn't be expected to be i.i.d. Some jokes may be broadly perceived to be funnier than others, and some users may be more critical than others. It may be desirable to transform the data to account

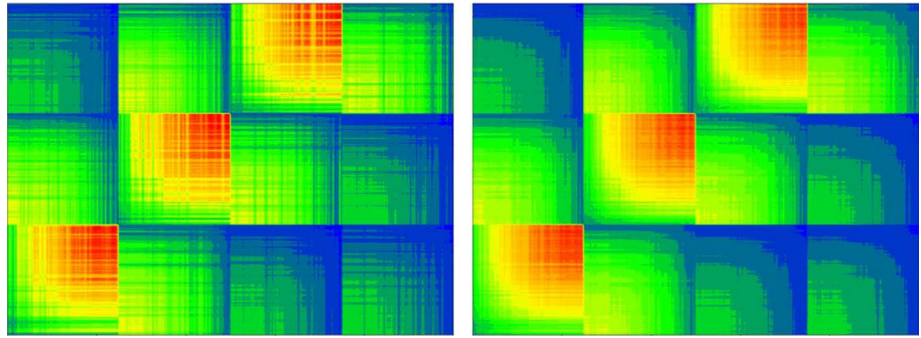

**Fig. 3** Estimates of 90% missing network when using the true community memberships

**Table 1** Comparing the performance of our model against alternatives on the simulated data

| Missing proportion | NMI of row community estimates | NMI column community estimates | Proposed method test set MSE | Proposed method with known communities test set MSE | NCF test set MSE | IBCF test set MSE | Funk SVD test set MSE |
|---|---|---|---|---|---|---|---|
| 0 | 1 | 1 | 1.87 (training set) | 1.87 (training set) | 3.0 (training set) | NA | NA |
| 0.5 | 1 | 1 | **6.44** | 6.44 | 9.91 | 55.42 | 5107.77 |
| 0.5 | 1 | 1 | **6.62** | 6.62 | 9.45 | 54.94 | 5114.18 |
| 0.5 | 1 | 1 | **5.96** | 5.96 | 15.23 | 53.70 | 5127.62 |
| 0.5 | 1 | 1 | 6.30 | 6.30 | **4.88** | 53.68 | 5099.26 |
| 0.5 | 1 | 1 | **6.37** | 6.37 | 12.14 | 57.24 | 5117.81 |
| 0.7 | 1 | 1 | **20.44** | 20.44 | 60.26 | 72.32 | 5110.85 |
| 0.7 | 1 | 1 | **18.12** | 18.12 | 48.36 | 69.59 | 5143.52 |
| 0.7 | 1 | 1 | **22.40** | 22.40 | 29.57 | 73.81 | 5124.30 |
| 0.7 | 1 | 1 | **21.92** | 21.92 | 33.42 | 73.54 | 5091.43 |
| 0.7 | 1 | 1 | 20.70 | 20.70 | **19.15** | 71.55 | 5162.01 |
| 0.8 | 0.98 | 1 | **52.36** | 51.16 | 278.42 | 125.56 | 5137.09 |
| 0.8 | 1 | 1 | **47.56** | 47.56 | 206.19 | 116.28 | 5103.40 |
| 0.8 | 0.96 | 1 | **58.87** | 52.61 | 160.97 | 127.31 | 5122.81 |
| 0.8 | 0.99 | 1 | **50.73** | 50.12 | 192.34 | 132.46 | 5086.41 |
| 0.8 | 0.99 | 1 | **50.43** | 49.55 | 83.82 | 117.93 | 5111.58 |
| 0.9 | 0.49 | 0 | 1667.46 | 152.13 | 2262.41 | 1532.01 | **336.04** |
| 0.9 | 0.50 | 0 | 1736.39 | 145.21 | 2004.16 | 1473.11 | **251.05** |
| 0.9 | 0.44 | 0.03 | 1129.04 | 169.02 | 1906.90 | 1567.24 | **311.70** |
| 0.9 | 0 | 0.45 | 1806.97 | 167.90 | 1891.07 | 1513.37 | **390.95** |
| 0.9 | 0 | 0.46 | 1851.84 | 159.07 | 2057.29 | 1466.07 | **267.55** |

Varying the proportion of hidden ratings, hiding 50%, 70%, 80%, and 90% of the data. Community detection performance is measured using the NMI between the recovered communities and the true communities. The best result for each run (based on test set MSE) is bolded. The gray column represents the test set performance of our estimation procedure if the true communities are known in advance for these effects, as it may be more informative to identify nodes with similar *relative* order of preferences. For this reason, when performing community detection only, we do not just use the observed matrix of ratings, but also the matrix after column centering, after row centering, after column normalizing, and after row normalizing. Column (row) centering means each column (row) will have a mean of 0, while column (row) normalizing means each column (row) will be demeaned and have a standard deviation of 1.

From the 74% dense dataset described above, we create 9 (non-independent) experiments. In the first 3, we keep 75% of the edges (both observed and missing) in the training set, and put the remaining 25% in the test set. Accounting for the initial network's density, the training sets have about 55.5% density. In the next 3, the train-test split is 50/50, so the training set is about 37% dense. The final 3 examples, the train test split is 25/75, with training set density about 18.5%. As the test sets are chosen randomly, edges are expected to appear in multiple test sets. For each of these 9 training sets, we conduct 3 fold cross validation. For every fold, we estimate communities based on the remaining 2 folds of the training set. Then, using the estimated communities and the original values (not including the test set or validation fold), we estimate the network, running 10 iterations to impute and stabilize the missing values. These estimates are compared to the edge weights in the validation fold. The transformation that yields the minimal average Normalized Mean Absolute Error (NMAE) is used for community detection for this experiment. Although MSE was the metric for the simulated data, we use NMAE here, following Goldberg et al. (2001). As the range of the data is $-10$ to 10, NMAE is MAE divided by 20. Finally, communities are estimated using the entire training set and the best performing transformation. Those communities are used to estimate the edge weights in the test set, iterating 10 times.

Table 2 contains detailed results for these experiments. Unlike the simulation dataset, true communities are unknown here, so NMI values can't be computed. Instead, we include the NCF performance on the training set, to demonstrate that the NCF approach is likely overfitting to the training set. All models perform worse with sparser training sets. Our model is best in the sparsest case, while approaches from recommenderlab achieve better results when more data is available. For this dataset, IBCF dominates all other algorithms besides our proposed model on these experiments. These results indicate that the proposed approach may not be well-suited to detecting subtler signals restricted to a smaller subset of users and items, leaving room for enhancement. Nonetheless, as this model can robustly recover large scale, coarse patterns in the data, there appears to be a sparsity regime where the proposed approach provides better than state-of-the-art performance. As real world datasets are often extremely sparse, the sparsest setting may most resemble realistic conditions, albeit without including additional features about users and items.

Figure 4 shows the results of our approach and NCF, with all networks and predictions reordered by the communities detected in the run depicted in the bottom row of Fig. 4, which is why that plot contains smooth contours. Each run detected different community structures, but a consistent ordering was chosen in order to depict broader trends throughout the different results. The ordering of the original dataset is not informative beyond the fact that certain jokes have many more ratings than other (the less rated jokes look like white strips in the top left plot in Fig. 4). This doesn't appear to be related to the content of the jokes, and splitting the data into training and test sets obviates some of this disparity. Though there is perhaps more fine grained prediction with more data, our method's predictions do not appear to change too much even in sparse settings. The

**Table 2** Comparing the performance of our model against alternatives on the Jester Dataset for users who rated exactly 74 jokes

| Approximate missing proportion | Transformation | Number of row communities detected | Number of column communities detected | NMAE for proposed model | NMAE for NCF | NMAE for NCF training set | NMAE for IBCF (average of 10) |
| --- | --- | --- | --- | --- | --- | --- | --- |
| 0.445 | None | 15 | 1 | 0.1760 | 0.1860 | 0.1312 | **0.1680** |
| 0.445 | None | 4 | 1 | 0.1737 | 0.1831 | 0.1298 | **0.1655** |
| 0.445 | Center Columns | 14 | 1 | 0.1739 | 0.1831 | 0.1296 | **0.1659** |
| 0.63 | None | 6 | 1 | 0.1766 | 0.1952 | 0.1254 | **0.1693** |
| 0.63 | None | 7 | 2 | 0.1769 | 0.1912 | 0.1268 | **0.1689** |
| 0.63 | None | 3 | 2 | 0.1758 | 0.1940 | 0.1233 | **0.1703** |
| 0.815 | None | 5 | 1 | **0.1804** | 0.2082 | 0.09602 | 0.1850 |
| 0.815 | None | 4 | 3 | **0.1812** | 0.2082 | 0.0963 | 0.1838 |
| 0.815 | None | 5 | 2 | **0.1796** | 0.2103 | 0.0961 | 0.1850 |
| 0.8125 (75% dense with further 75% hidden) | No transformation, warm start for the column communities | 3 | 4 | **0.1762** | 0.1960 | 0.1028 | 0.1889 |

Varying the proportion of ratings which are hidden, hiding ∼ 25%, ∼ 50%, and ∼ 75% of the 74% dense network. As true communities are unknown, the number of recovered row and column communities are reported by the best performing transformation of the data according to training set cross validated NMAE. The best result for each run based on test set NMAE is bolded. The gray column represents training set performance of the NCF, indicating alternative hyperparameters are unlikely to drastically improve NCF test set performance. The final row contains results for a different dataset, users who rated exactly 75 jokes right two columns of Fig. 4 show some NCF estimates appear to be more extreme. In contrast to our model and NCF, recommenderlab only makes predictions on the test set, so plots would be misleading as they would frequently include observations directly from the training set.

Based on the resulting performance of the recovered communities on different folds of the training set, we can determine which transformation appears to find meaningful sets of communities. In practice, using the original dataset led to better performance than any transformation in 8 of 9 experiments. As our community detection algorithm is slow, cross validation took far longer than any other aspect of the modeling pipeline. Our recommendation is therefore to skip cross validation and simply perform community detection and estimation on the original training set.

Finally, to ensure that this performance is not limited to the dataset of users who rated exactly 74 jokes, we attempted a similar analysis with the dataset of users who rated exactly 75 jokes. Though the jokes are the same, this is a different dataset, as all users are previously unobserved. We included only 25% of data in the training set, leaving an approximately 18.75% dense network as the training set. Starting with a previously recovered set of item (joke) communities from the eighth run in Table 2, but initializing all users as singletons, we re-ran community detection without cross validation, and ran our estimation procedure. In this case of "transfer learning," our model again outperformed other approaches, indicating the model's success in this sparsity regime is not simply a quirk of the previous dataset.

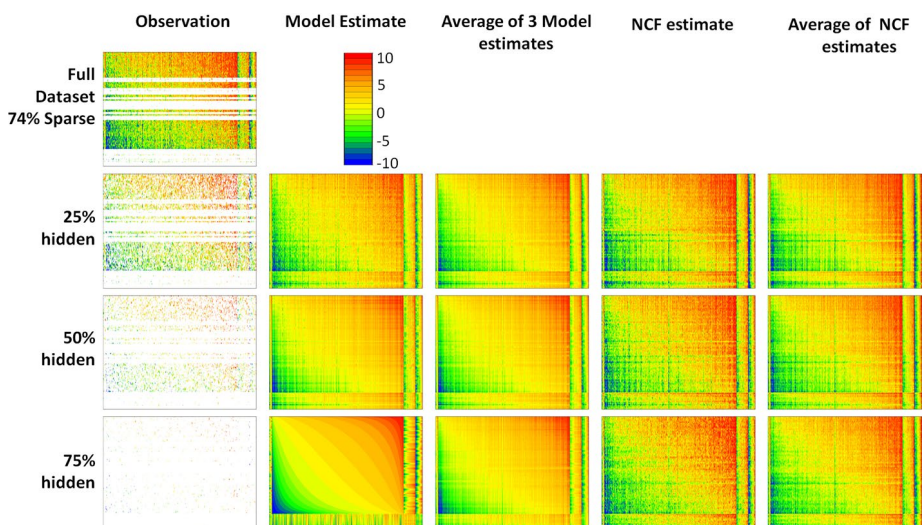

**Fig. 4** By column from left to right: Observed networks, our model's estimates, the average of 3 model estimates, NCF estimates, and the average of 3 NCF estimates. For legibility purposes, all networks are reordered according to the estimated communities as detected by the eighth row in Table 2, with 4 row communities and 3 column communities, which is depicted in the bottom plot in the second column from the left

### Future work

While our proposed methodology provides promising results, there is room for improvement. In real world, online recommender systems, additional information is often known about users or items which may be leveraged to improve recommendations, especially for new users or items. One option would be to generalize the model discussed here, but it might be simpler to adapt the present notion of communities into a NCF, as was done with (Tsitsulin et al. 2023), which has communities based on modularity. That option may combine the functional fitting strengths of neural networks while restricting the fit within estimated subnetworks. However, the community detection process is too slow for industrial scale. Faster approaches are necessary, likely requiring developments in both theory and algorithms. It is difficult to upgrade from a "good" clustering to an optimal or near optimal one because the measure $L$ is based on node-community correlations, and because poor user clusterings can cascade into poor item clusterings or vice versa.

Though cross validation was used to try to find the best transformation of the data, it ultimately proved unhelpful. However, cross validation could be used to choose model complexity. One interpretation of the slower rate of performance decay for our model is that our model is able to find coarse, globally consistent patterns with few data points. However, given more data, algorithms like IBCF may be able to pick up cross-cutting patterns that are too fine for our method to detect. Some distinct communities may be incorrectly merged based on similar large scale dynamics, displaying a kind of resolution limit. Given our results, it may be useful to run the estimation pipeline, and subsequently run it on the matrix consisting of residuals of the initial estimates, allowing for more complicated behaviors. Cross validation can be used to determine the appropriate model complexity. Ideally, this would encourage a model to capture only simpler, global patterns when data is sparse, while using more locally appropriate information when that signal can be detected using a richer dataset, achieving excellent performance in both regimes.

## Appendix 1: Algorithms

This appendix contains supplementary algorithms for community detection for partially observed dense weighted bipartite networks as described in "Community detection algorithm" section. The overarching algorithm displayed as Algorithm 1 is relatively simple, though the implementation is technically complex.

The first two algorithms contained in this appendix are used as helper functions for Algorithm 2, used to combine distinct communities into a single entity.

**Algorithm 5** attemptMerge for bipartite networks

---

**Procedure** attemptMergeBipartite($W$, *communityAggregate*, *relevantLabels*, *relevantLabelOrder*, *otherLabels*)

1    newLabels = relevantLabels
2    newOrder = relevantLabelOrder
3    A = newOrder(1)
4    B = $\underset{B' \neq A}{\operatorname{argmax}}\, corr([\text{communityAggregate(A)}], [\text{communityAggregate(B')}])$

```
// [communityAggregate(A)] represents the Ath column in the
   communityAggregate matrix (either rowCommunityAggregate or
   columnCommunityAggregate). Correlation is the "pairwise
   complete" correlation here and everywhere else in this
   document. This algorithm seeks to combine community ''A''
   with the community whose edge weights are most correlated
   with A's
```

5    mergedLabels = newLabels
6    mergedLabels(mergedLabels = B) = A
7    mergedOrder = newOrder
8    mergedOrder(mergedOrder = B) = A

```
// The above four lines will tentatively merge the communities
   represented by columns A and B to see if it improves the
   measure L below
```

9    **if** *calculateMeasureL(W, mergedLabels, otherLabels)* $\geq$ *calculateMeasureL(W, newLabels, otherLabels)* **then**
10      newLabels = mergedLabels
11      newOrder = mergedOrder
12      [communityAggregate(A)] += [communityAggregate(B)]
13      Remove [communityAggregate(B)] column from communityAggregate

```
// If the measure L is improvedby combining the
   communities, update both in the community labels and in
   the relevant communityAggregate matrix
```

     **end**
14    **return** newLabels, communityAggregate, newOrder

**Algorithm 6** Sweep for bipartite networks

---
**Procedure** `sweepBipartite`(*W, communityAggregate, relevantLabels, otherLabels*)

1. correlationOrder = sort non-matching label pairs by decreasing correlations of columns in communityAggregate with each row containing 2 communities
2. **while** *length(correlationOrder)* > 0 **do**
3.     newLabels = relevantLabels
4.     A = correlationOrder(1, 1)   // First community in first row of correlationOrder
5.     B = correlationOrder(1, 2)   // Second community in first row of correlationOrder
6.     mergedLabels = newLabels
7.     mergedLabels(mergedLabels = B) = A
8.     **if** *calculateMeasureL(W, mergedLabels, otherLabels)* ≥ *calculateMeasureL(W, newLabels, otherLabels)* **then**
9.         newLabels = mergedLabels
10.         [communityAggregate(A)] += [communityAggregate(B)]
11.         Remove [communityAggregate(B)] column from communityAggregate
12.         dequeueAll(correlationOrder)
       **else**
13.         dequeue(correlationOrder(1))
        **end**
   **end**
14. **return** newLabels, communityAggregate
---

The remaining algorithms operationalize heuristics for reclassifying nodes. The first step is identifying nodes which may be misclassified. Going back to "Community detection" section, we expect a good estimated clustering to have large positive $C_{\widehat{ij}}$ values everywhere. Algorithms 7 and 8 are built on the insight that if there is a negative $C_{\widehat{ij}}$ value, then a row (or column) node's edges are anticorrelated with its assigned community's edges, with respect to some column (row) community, which may mean the row node's comunity assignment is misclassified. However, if all the negative row $C_{\widehat{ij}}$ values are due to one particular column community *j*, then perhaps that whole estimated column community is the root of the problem. Algorithms 7 and 8 handle this issue somewhat differently. If all row node issues are due to only one column community, Algorithm 7 does not consider these row nodes to be misclassified, while Algorithm 8 still does. Nevertheless, the purpose of each is to identify those problematic nodes and remove their current community labels. Algorithm 7 sends any potentially misclassified node to its own new singleton community. Algorithm 8 will remove problematic nodes from their current communities, but will take any row nodes that came from the same community that also have negative $C_{\widehat{ij}}$ values with respect to the exact same row communities, and place them together into a newly formed community. Especially in the case where many edges are missing, having communities with multiple nodes may make future steps more robust. However, because having exactly two nodes in a community can make it very easy for correlations to be very high, Algorithm 8 also breaks up any community of exactly two nodes into two different communities. The reason for including both of these algorithms is clustering may be unstable in myriad ways, so these approaches take different paths, giving more options to find marginally better, or even much better clusterings. Aggomeration is run on both of these results to see if the newly "pure" communities can be recombined, but there are also two more algorithms for recombination.

**Algorithm 7** removeNodesA

---

**Procedure** removeNodesA($W$, $rowLabels$, $columnLabels$)

1. rowCorrMat = $n_r \times \widehat{K_c}$ matrix where entry $(u, j)$ contains $C_{\hat{i}\hat{j}}(u)$ where $u \in \hat{i}$
2. columnCorrMat = $n_c \times \widehat{K_r}$ matrix where entry $(v, i)$ contains $C_{\hat{j}\hat{i}}(v)$ where $v \in \hat{j}$

   ```
   // n_r and n_c are the number of rows and columns in the
      network, respectively. rowCorrMat_{u,j} shows correlation
      between the edge weights connecting row node u to estimated
      column community j and the total degree of column nodes in
      estimated column community j with respect to estimated row
      community i. columnCorrMat_{v,i} shows similar quantities for
      column node v, estimated row community i, and estimated
      column community j. Any negative correlations may indicate
      a clustering issue. This algorithm moves each problematic
      node to a new singleton community
   ```

3. rowWrongNodes = {}; columnWrongNodes = {}
4. correspondingColumns = {}; correspondingRows = {}   // Every top level for loop or if statement done for row nodes will be done correspondingly immediately thereafter for column nodes. From here on, for simplicity, only the row node steps are shown.
5. **for** $u$ in 1 to $n_r$ **do**
      **if** $min(rowCorrMat_i) < 0$ **or** $max(rowCorrMat_i) = 0$ **then**
         queue(rowWrongNodes, u)
         queue(correspondingColumns, which($rowCorrMat_i$)<0)
         // Mark problematic nodes and where they cause issues
      **end**
   **end**
6. **if** $max(correspondingColumns) = min(correspondingColumns)$ **and** $length(correspondingColumns) \geq |columnLabels = min(correspondingColumns)|$ **then**
      rowWrongNodes = {}; queue(columnWrongNodes, $\{v : columnLabels(v) = min(correspondingColumns)\}$ )
   **end**
   // if all problematic row nodes are due to one column community, and there are more marked row nodes than members of that column community, mark that whole column community as problematic, and unmark all row nodes
7. **for** $u$ in $rowWrongNodes$ **do**
      $\widehat{K_r}$++; rowLabels(u) = $\widehat{K_r}$ // move marked nodes to own community
   **end**
8. L = calculateMeasureL(W, rowlabels, columnLabels)
9. **return** rowLabels, columnLabels, rowWrongNodes, rowColumnNodes, L

**Algorithm 8** removeNodesB

---

**Procedure** removeNodesB(*W*, *rowLabels*, *columnLabels*)

1  // $n_r$, $n_c$, *rowCorrMat*, and *colCorrMat* are as in earlier algorithms. Negative values may indicate misclustering. This algorithm moves problematic nodes to new communities using patterns of negative correlations
2  rowWrongNodes = {}; columnWrongNodes = {}
3  correspondingColumns = {}; correspondingRows = {}; // Every top level for loop or if statement done for row nodes will be done correspondingly immediately thereafter for column nodes. From here on, for simplicity, only the row node steps are shown.
4  **for** *u in 1 to $n_r$* **do**
    **if** $min(rowCorrMat_i)<0$ **or** $max(rowCorrMat_i)= 0$ **or** $|rowLabels = rowLabels(u)| = 2$ **then**
        queue(rowWrongNodes, u)// same step as in above algorithm
        queue(correspondingColumns, which($rowCorrMat_i$) < 0)
    **end**
  **end**
5  **if** *max(correspondingColumns) = min(correspondingColumns)* **and** *length(correspondingColumns)$\geq$ |columnLabels = min(correspondingColumns)|* **then**
    queue(columnWrongNodes, $\{v : columnLabels(v) = min(correspondingColumns)\}$ ) // if all problematic row nodes are due to one column community, and there are more marked row nodes than members of that column community, mark that whole column community as problematic
  **end**
  additionalRowCommunities = 0
6  **for** *i in 1 to $\widehat{K_r}$* **do**
    binaryString = {}
    **for** $\{u : \{rowLabels(u) = i\} \cap rowWrongNodes\}$ **do**
        binaryString(u) = binaryToInteger($rowCorrMat_u < 0$) // nodes coming from same community that have same pattern of negatives are moved to same new community
    **end**
    **for** *val in unique(binaryString)* **do**
        additionalRowCommunities++
        rowLabels({u: binaryString(u) = val}) = $\widehat{K_r}$+additionalRowCommunities
    **end**
  **end**
7  L = calculateMeasureL(W, rowlabels, columnLabels)
  **return** rowLabels, columnLabels, rowWrongNodes, rowColumnNodes, L

---

Algorithm 9 is meant only to be run on the result of Algorithm 7, seeking to place those newly removed singleton communities into pre-existing communities. Algorithm 10 is a bit more radical, in that it uses the same criteria to move communities into other communities, but it will even combine pre-existing communities. Furthermore, both algorithms have a toggle called banRemain, which, if turned on, prevents nodes that have just been removed from their previous community from rejoining that community. Algorithm 9 is run both with and without banRemain on the reuslt of Algorithm 7, and Algorithm 8 is run both with and without banRemain on the result of 8, as well as without banRemain on the result of 7. Agglomeration is subsequently run on all these results. All told, that will create 14 clustering options to be compared with the current result.

**Algorithm 9** regoupNodesA

---

**Procedure** `regoupNodesA(`$W, rowLabels, columnLabels, rowWrongNodes,$
$rowColumnNodes, banRemain, previousRowLabels, previousColumnLabels$`)`

1. **for** $i$ in $max(previousRowLabels)+1$ to $max(rowLabels)$ **do**
2.     **for** $k$ in $1$ to $max(previousRowLabels)$ **do**
   
   $$\text{rowCommunityvalues}(i, k) = \sum_{j=1}^{\widehat{K_c}} max(0, n_k - 2) \times corr\big(W(i, v \in j), rowCommunityAggregate(k, v \in j)\big)$$
   
       **end**
   
   **end**
   
   // repeat above for loop for columns

3. **for** $i$ in $max(previousRowLabels)+1$ to $max(rowLabels)$ **do**
   
       **if** $banRemain = 0$ **then**
   
           $\text{rowLabels}(\{u : rowLabels = i\}) = \underset{i' \in \{1,...,max(previousRowLabels)\}}{\arg\max} (rowCommunityvalues_i)$
   
       **end**
   
       **else**
   
           $\text{rowLabels}(\{u : rowLabels = i\} \cap \{u \notin rowWrongNodes\}) = \underset{i' \in \{1,...,max(previousRowLabels)\}}{\arg\max} (rowCommunityvalues_i)$
   
           $\text{rowLabels}(\{u : rowLabels = i\} \cap \{u \in rowWrongNodes\}) = \underset{i' \in \{1,...,max(previousRowLabels)\} \setminus \{previousNodeLabels(u)\}}{\arg\max} (rowCommunityvalues_i)$
   
       **end**
   
   **end**
   
   // repeat above for loop for columns

4. $L = calculateMeasureL(W, rowlabels, columnLabels)$
5. **return** $rowLabels, columnLabels, L$

**Algorithm 10** regroupNodesB

**Procedure** regoupNodesB(*W, rowLabels, columnLabels, rowWrongNodes, rowColumnNodes, banRemain, previousRowLabels, previousColumnLabels*)

1    **for** *i in 1 to max(rowLabels)* **do**
2      **for** *k in 1 to max(rowLabels)* **do**
       rowCommunityvalues(i, k) =
$$\sum_{j=1}^{\widehat{K_c}} max(0, n_k - 2) \times corr\big(rowCommunityAggregate(i, v \in j), rowCommunityAggregate(k, v \in j)\big)$$
     **end**
   **end**
   // repeat above for loop for columns
3    **for** *i in 1 to max(rowLabels)* **do**
     **if** *banRemain = 0* **then**
       rowLabels($\{u : rowLabels = i\}$) = $\underset{i' \in \{1,...,max(rowLabels)\}}{\mathrm{argmax}} (rowCommunityvalues_i)$
     **end**
     **else**
       rowLabels($\{u : rowLabels = i\} \cap \{u \notin rowWrongNodes\}$) = $\underset{i' \in \{1,...,max(rowLabels)\}}{\mathrm{argmax}} (rowCommunityvalues_i)$
       rowLabels($\{u : rowLabels = i\} \cap \{u \in rowWrongNodes\}$) = $\underset{i' \in \{1,...,max(rowLabels)\} \setminus \{previousNodeLabels(u)\}}{\mathrm{argmax}} (rowCommunityvalues_i)$
     **end**
   **end**
   // repeat above for loop for columns
4    L = calculateMeasureL(W, rowlabels, columnLabels)
5    **return** rowLabels, columnLabels, L

**Abbreviations**
GCN    Graph convolutional network
GNN    Graph neural network
IBCF    Item-based collaborative filtering
MCAR    Missing completely at random
NCF    Neural Collaborative Filtering
NMAE    Normalized mean absolute error
NMI    Normalized mutual information
NSM    Nonlinear sociability model
SVD    Singular value decomposition
UBCF    User-based collaborative filtering

**Author contributions**
BL created and implemented the algorithms. VP assisted in designing the approach and interpreting results. All authors contributed to writing and editing the manuscript, and read and approved the final manuscript.

**Funding**
VP wishes to acknowledge support for this work from NSF DMS-2113662 and NSF RTG grant DMS-2134107.

**Availability of data and materials**
The Jester dataset analysed during the current study are available at https://eigentaste.berkeley.edu/dataset/jester_dataset_1_1.zip (Jester Dataset xxx) The full dataset generated for the simulations in Fig. 2 is available at https://raw.githubusercontent.com/bleinwand/bleinwand.github.io/master/recommender_system_canonical_example_matrix.csv.

## Declarations

**Competing interests**
The authors declare that they have no Conflict of interest.

## Publisher's Note